\documentclass[aps,prc,twocolumn,superscriptaddress]{revtex4-1}

\usepackage{graphics}

\begin{document}


\title{Inverse-kinematics proton scattering from $^{42,44}$S,
  $^{41,43}$P and the collapse of the $N=28$ major shell closure}



\author{L. A. Riley}
\affiliation{Department of Physics and Astronomy, Ursinus College,
  Collegeville, PA 19426, USA}

\author{D. Bazin} \affiliation{National Superconducting Cyclotron
  Laboratory, Michigan State University, East Lansing, MI, 48824, USA}
\affiliation{Department of Physics and Astronomy, Michigan State
  University, East Lansing, MI, 48824, USA}

\author{J. Belarge} \altaffiliation{J. Belarge is currently a MIT
  Lincoln Laboratory employee. No Laboratory funding or resources were
  used to produce the results/findings reported in this article.}

\affiliation{National Superconducting Cyclotron
  Laboratory, Michigan State University, East Lansing, MI, 48824, USA}
\affiliation{Department of Physics and Astronomy, Michigan State
  University, East Lansing, MI, 48824, USA}

\author{P. C. Bender} \affiliation{National Superconducting Cyclotron
  Laboratory, Michigan State University, East Lansing, MI, 48824, USA}

\author{B. A. Brown} \affiliation{National Superconducting Cyclotron
  Laboratory, Michigan State University, East Lansing, MI, 48824, USA}
\affiliation{Department of Physics and Astronomy, Michigan State
  University, East Lansing, MI, 48824, USA}

\author{P. D. Cottle} \affiliation{Department of Physics, Florida
  State University, Tallahassee, FL 32306, USA}

\author{B. Elman} \affiliation{National Superconducting Cyclotron
  Laboratory, Michigan State University, East Lansing, MI, 48824, USA}
\affiliation{Department of Physics and Astronomy, Michigan State
  University, East Lansing, MI, 48824, USA}

\author{A. Gade} \affiliation{National Superconducting Cyclotron
  Laboratory, Michigan State University, East Lansing, MI, 48824, USA}
\affiliation{Department of Physics and Astronomy, Michigan State
  University, East Lansing, MI, 48824, USA}

\author{S. D. Gregory} \affiliation{Department of Physics and
  Astronomy, Ursinus College, Collegeville, PA 19426, USA}

\author{E. B. Haldeman} \affiliation{Department of Physics and
  Astronomy, Ursinus College, Collegeville, PA 19426, USA}

\author{K. W. Kemper} \affiliation{Department of Physics, Florida
  State University, Tallahassee, FL 32306, USA}

\author{B. R. Klybor} \affiliation{Department of Physics and
  Astronomy, Ursinus College, Collegeville, PA 19426, USA}

\author{M. A. Liggett} \affiliation{Department of Physics and
  Astronomy, Ursinus College, Collegeville, PA 19426, USA}

\author{S. Lipschutz} \affiliation{National Superconducting Cyclotron
  Laboratory, Michigan State University, East Lansing, MI, 48824, USA}
\affiliation{Department of Physics and Astronomy, Michigan State
  University, East Lansing, MI, 48824, USA}

\author{B. Longfellow} \affiliation{National Superconducting Cyclotron
  Laboratory, Michigan State University, East Lansing, MI, 48824, USA}
\affiliation{Department of Physics and Astronomy, Michigan State
  University, East Lansing, MI, 48824, USA}

\author{E. Lunderberg} \affiliation{National Superconducting Cyclotron
  Laboratory, Michigan State University, East Lansing, MI, 48824, USA}
\affiliation{Department of Physics and Astronomy, Michigan State
  University, East Lansing, MI, 48824, USA}

\author{T. Mijatovic} \affiliation{National Superconducting Cyclotron
  Laboratory, Michigan State University, East Lansing, MI, 48824, USA}

\author{J. Pereira} \affiliation{National Superconducting Cyclotron
  Laboratory, Michigan State University, East Lansing, MI, 48824, USA}
\affiliation{Department of Physics and Astronomy, Michigan State
  University, East Lansing, MI, 48824, USA}

\author{L. M. Skiles} \affiliation{Department of Physics and
  Astronomy, Ursinus College, Collegeville, PA 19426, USA}

\author{R. Titus} \affiliation{National Superconducting Cyclotron
  Laboratory, Michigan State University, East Lansing, MI, 48824, USA}
\affiliation{Department of Physics and Astronomy, Michigan State
  University, East Lansing, MI, 48824, USA}

\author{A. Volya} \affiliation{Department of Physics, Florida State
  University, Tallahassee, FL 32306, USA}

\author{D. Weisshaar} \affiliation{National Superconducting Cyclotron
  Laboratory, Michigan State University, East Lansing, MI, 48824, USA}

\author{J.C. Zamora} \affiliation{National Superconducting Cyclotron
  Laboratory, Michigan State University, East Lansing, MI, 48824, USA}

\author{R. G. T. Zegers} \affiliation{National Superconducting
  Cyclotron Laboratory, Michigan State University, East Lansing, MI,
  48824, USA} \affiliation{Department of Physics and Astronomy,
  Michigan State University, East Lansing, MI, 48824, USA}
\affiliation{Joint Institute for Nuclear Astrophysics - Center for the
  Evolution of the Elements, Michigan State University, East Lansing,
  MI 48824, USA}

\date{\today}

\begin{abstract}
Excited states of the neutron-rich isotopes $^{42,44}$S and
$^{41,43}$P have been studied via inverse-kinematics proton scattering
from a liquid hydrogen target, using the GRETINA $\gamma$-ray tracking
array to extract inelastic scattering cross sections. Deformation
lengths of the $2^+_1$ excitations in $^{42,44}$S have been determined
and, when combined with deformation lengths determined with
electromagnetic probes, yield the ratio of neutron-to-proton matrix
elements $M_n/M_p$ for the $2^+_1$ excitations in these nuclei. The
present results for $^{41,43}$P$(p,p')$ are used to compare two shell
model interactions, SDPF-U and SDPF-MU.  As in a recent study of
$^{42}$Si, the present results on $^{41,43}$P favor the SDPF-MU
interaction.
\end{abstract}

\pacs{}

\maketitle


\section{Introduction}

One of the highest scientific priorities for nuclear structure
physicists during the last few decades has been to determine the
behavior of the major neutron shell closure at $N=28$ and to
understand the mechanism underlying its collapse in neutron-rich
nuclei near $^{42}$Si, which is close to the neutron drip line.  This
shell closure is strongly defined in the stable $N=28$ isotone
$^{48}$Ca, but appears to narrow and then collapse as protons are
removed.  The energy of the $2^+_1$ state decreases from 3832 keV in
$^{48}$Ca \cite{Bu06} to 1329~keV in the radioactive nucleus $^{44}$S
\cite{Che11,Par17} and then to 742~keV in $^{42}$Si \cite{Ch16}.  In fact,
the $2^+_1$ state energy in $^{42}$Si is lower than it is in the
$N=26$ isotope $^{40}$Si (986~keV \cite{Ch17}) so that 
the most recognizable signature of a major shell closure -- a
significant increase in the energy of the $2^+_1$ state -- has
disappeared entirely in the Si isotopes at $N=28$.

In the present work, we report results of inelastic proton scattering
studies of the radioactive $N=28$ isotones $^{44}$S and $^{43}$P and
the $N=26$ isotones $^{42}$S and $^{41}$P performed in inverse
kinematics with a liquid hydrogen target to uncover several new
aspects of the behavior of nuclei in the vicinity of $^{42}$Si.  In
$^{42, 44}$S, we are able to compare the results of the present inelastic
scattering measurement of the $2^+_1$ states to previous Coulomb
excitation measurements of the same transitions to determine whether
the excitations of these states are isoscalar.  In addition, we use the
results of the $^{41,43}$P$(p,p')$ measurements to compare the SDPF-U
and SDPF-MU shell model interactions, as was done in a recent study
of $^{42}$Si \cite{Gad19}.

\section{Experimental details}

The experiment was performed at the Coupled-Cyclotron Facility of the
National Superconducting Cyclotron Laboratory at Michigan State
University (NSCL)~\cite{NSCLFRIB}. The secondary beams were produced
by fragmentation of a 140~MeV/nucleon $^{48}$Ca primary beam in a
1222~mg/cm$^2$ $^9$Be production target and separated by the A1900
fragment separator~\cite{A1900}. The momentum acceptance of the A1900
was set to 2\%. A 300~mg/cm$^2$ aluminum achromatic wedge was used to
further separate the secondary beams by $Z$.  The beams of interest in
the present work were produced with two magnet settings of the A1900
and are summarized in Table~\ref{tab:beams}.

Secondary beam particles were identified upstream of the reaction
target by times of flight from the A1900 extended focal plane and the
object position of the S800 spectrograph~\cite{S800}. A
scintillator in the focal plane of the S800 was used to stop both
timing measurements.  The beam then passed through the NSCL/Ursinus
College Liquid Hydrogen Target, based on the design of Ryuto
\textit{et al.}~\cite{Ryu05}. The target was installed at the target
position of the S800. Outgoing beam particles
were identified by energy loss in the S800 ionization chamber and time of
flight. The reaction kinematics were such that all four
beams, over the full range of possible projectile kinetic energies
within the target, were scattered into laboratory angles below
$2^\circ$, falling entirely within the $7^\circ \times 10^\circ$
angular acceptance of the S800. 
The GRETINA $\gamma$-ray tracking array~\cite{GRETINA,
  GRETINA2} was centered on the target. Eight modules housing four
36-fold segmented high-purity germanium crystals were installed on one
of the GRETINA mounting hemispheres to accommodate the target. Two
modules were centered at 58$^\circ$, four at 90$^\circ$, and two at
122$^\circ$ with respect to the beam axis.

\begin{table}
  \caption{\label{tab:beams} Secondary beam properties and yields.}
  \begin{ruledtabular}
    \begin{tabular}{ccccc}
      Secondary & Purity & Mid-Target Energy & Mid-Target & Total
      \\ Beam & (\%) & (MeV/nucleon) & $v/c$ & Particles \\\hline
      $^{42}$S & 2  & 62.5 & 0.349 & $9.46 \times 10^5$ \\
      $^{41}$P & 30 & 57.7 & 0.336 & $1.37 \times 10^7$ \\ \hline
      $^{44}$S & 32 & 70.2 & 0.368 & $5.14 \times 10^6$ \\
      $^{43}$P & 9  & 64.7 & 0.354 & $1.37 \times 10^6$ \\
    \end{tabular}
  \end{ruledtabular}
\end{table}

The liquid hydrogen was contained by a cylindrical aluminum target
cell with 125~$\mu$m Kapton entrance and exit windows, mounted on a
cryocooler. The nominal target thickness was 30~mm.  The target cell
and cryocooler were surrounded by a 1~mm thick aluminum radiation
shield with entrance and exit windows covered by 5~$\mu$m aluminized
Mylar foil. The temperature and pressure of the target cell at
17.00(25)~K and 880(10)~Torr were monitored throughout the
experiment. The variations in the temperature and pressure of the
target cell corresponded to a 0.3~\% uncertainty in target density.

The pressure difference across the Kapton entrance and exit windows
caused them to bulge outward.  The resulting additional target
thickness was determined by fitting \textsc{geant4}~\cite{Geant4}
simulations of the beam particles traversing the target to the
measured kinetic energy distribution of the outgoing beam particles.
Before the liquid hydrogen target was filled, the kinetic energy
spectra of the secondary beams passing through the empty target cell
were measured.  Simulations of the beams passing through the full
target were run in which initial beam energies were drawn from
these measured empty-cell kinetic energy distributions. The thickness
of the outward bulge of the Kapton entrance and exit windows was
varied in the simulations of each beam, and the resulting outgoing
kinetic energy distributions were were fit by a simple scaling of the
simulated spectra.

\begin{figure*}
  \scalebox{0.57}{ \includegraphics{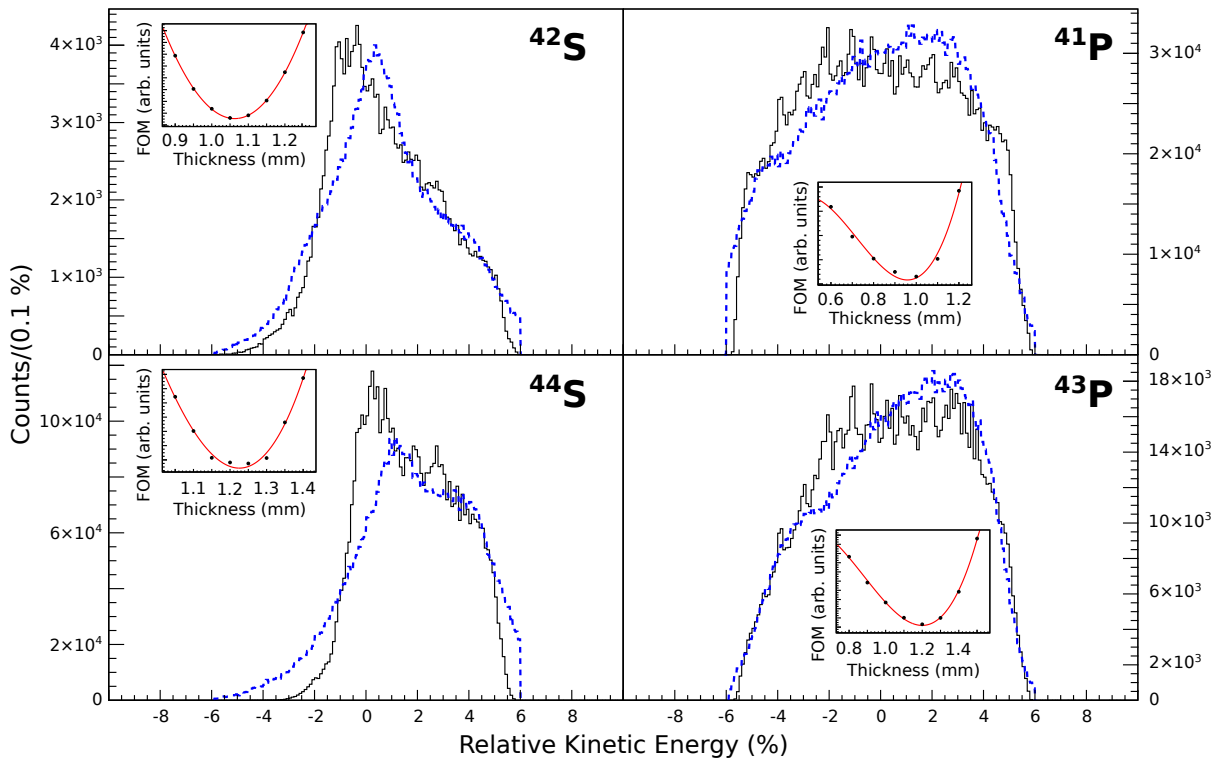} }
  \caption{\label{fig:bulge} (Color online)
    Relative kinetic energy spectra of the beams measured downstream 
    of the target in the S800. The dashed spectra are the
    \textsc{geant4} fits described in the text. The insets are plots
    of the figure of merit from log-likelihood fits of the simulated
    beam particles vs. the thickness of the outward bulge of the
    Kapton entrance and exit windows of the target.}
\end{figure*}

Measured kinetic energy spectra of the beams after traversing the full
target, relative to the kinetic energy corresponding to the center of
the S800 momentum acceptance, are shown in Fig.~\ref{fig:bulge}. The
dashed spectra in the four main panels of Fig.~\ref{fig:bulge} are
simulated spectra assuming the target bulge thickness 
giving the best fit to the measured spectra.  The
insets show the figure of merit from the log-likelihood fitting
procedure plotted vs. the simulated window bulge thickness.  This
process yielded best-fit bulge thicknesses of 1.06~mm and 0.96 mm for
the $^{42}$S and $^{41}$P beams and 1.20~mm and 1.22~mm for the
$^{44}$S and $^{43}$P beams. The statistical uncertainties in each of
these results, corresponding to the minimum figure of merit + 1, are
on the order of $10^{-3}$~mm. We attribute the discrepancies between
the best-fit simulations and the measured spectra, as well as the larger
$\approx 0.1$~mm observed variation among the best-fit bulge
thicknesses, to variations in the momentum distributions of the
incoming beams during the experiment and differences in the transverse
positions of the two secondary beams on the (curved) target, leading
to different effective target thicknesses. To determine the areal
density of the target for use in cross section calculations, we assume
a bulge thickness of 1.09(13)~mm, encompassing the full range of these
results, yielding an areal density of 240(2)~mg/cm$^2$. The mid-target
beam energies and average beam velocities given in
Table~\ref{tab:beams} were also determined using these simulations.

\section{Analysis and results}

\begin{figure}
  \scalebox{0.6}{ \includegraphics{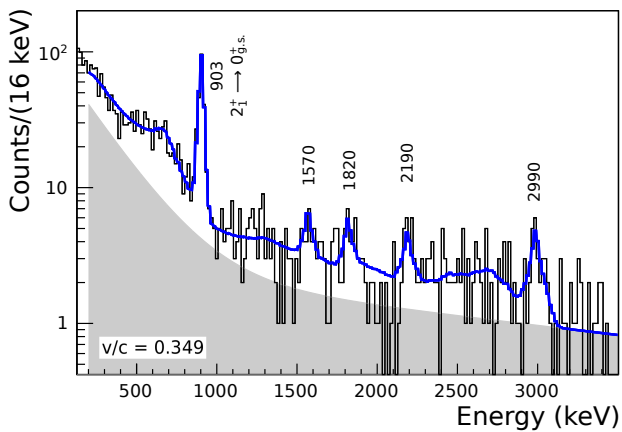} }
  \caption{\label{fig:s42_spectrum} (Color online) Projectile-frame
    spectrum of $^{42}$S measured via inverse-kinematics proton
    scattering.}
\end{figure}

\begin{figure}
  \scalebox{0.6}{ \includegraphics{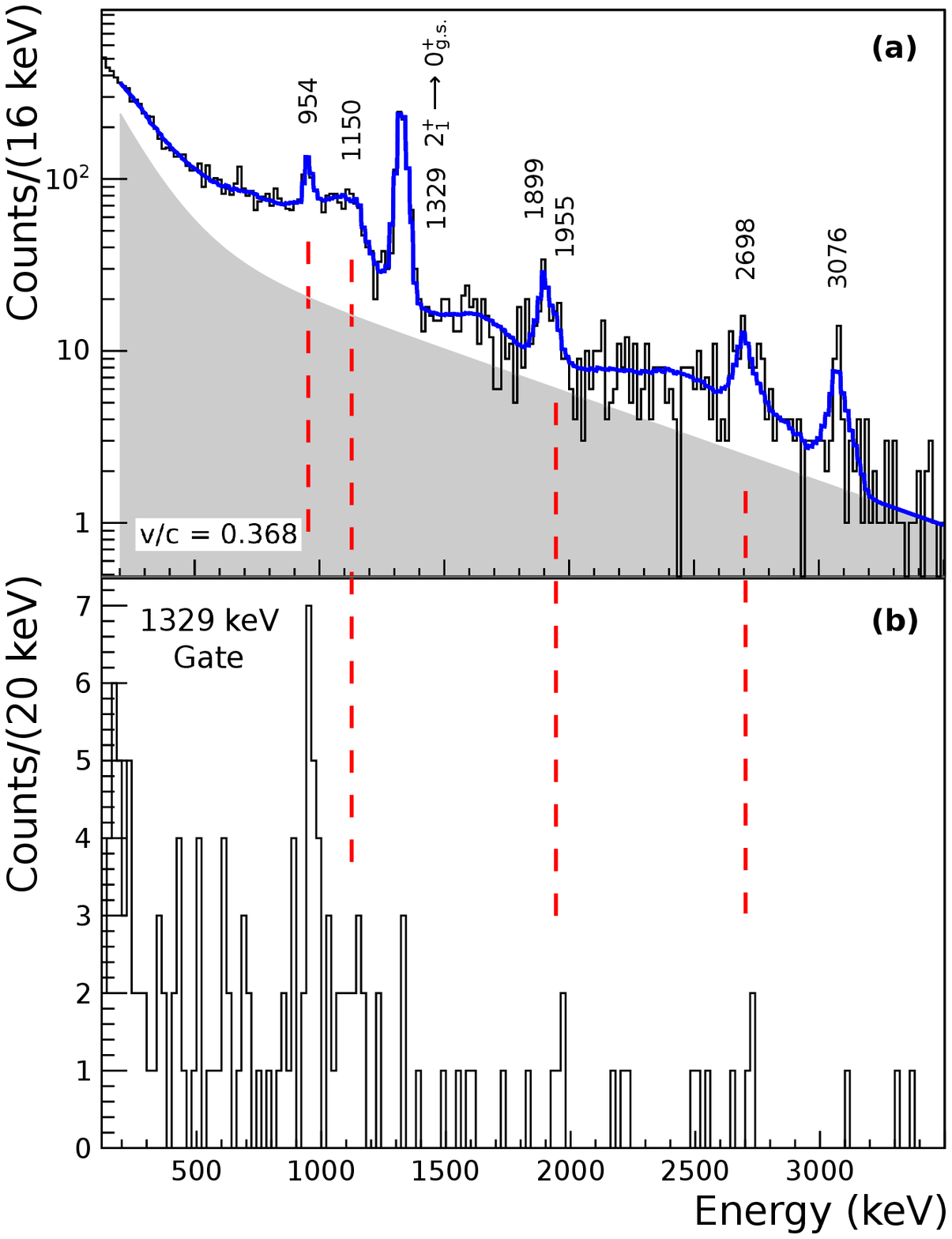} }
  \caption{\label{fig:s44_spectra} (Color online) (a) Projectile-frame
    spectrum of $^{44}$S measured via inverse-kinematics proton
    scattering. (b) Proton-scattering spectrum gated on the 1329~keV
    $\gamma$ ray.}
\end{figure}

\begin{figure} 
  \scalebox{0.6}{ \includegraphics{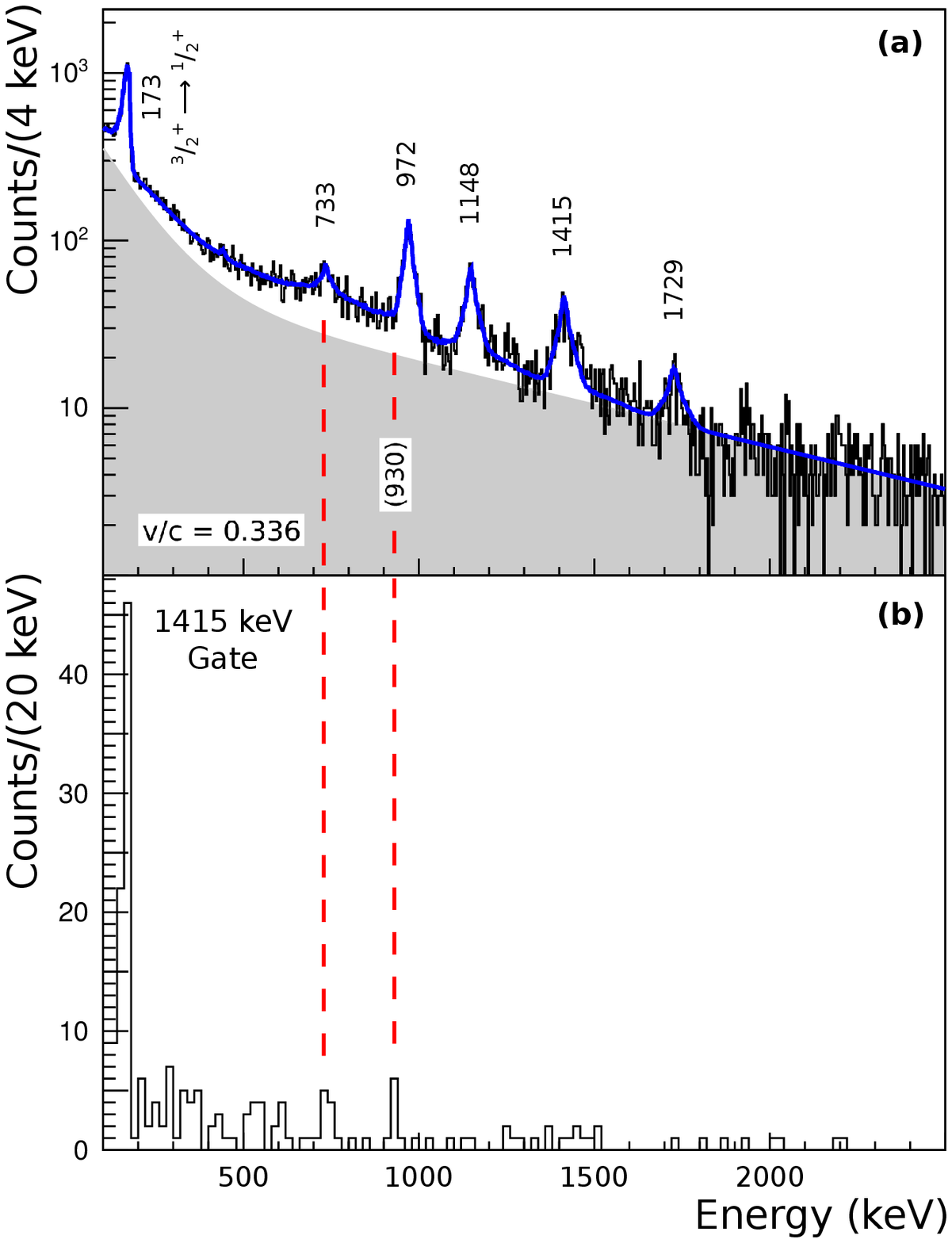} }
  \caption{\label{fig:p41_spectrum} (Color online) (a)
    Projectile-frame spectrum of $^{41}$P measured via
    inverse-kinematics proton scattering. (b) Proton-scattering
    spectrum gated on the 1416~keV $\gamma$ ray.}
\end{figure}

\begin{figure} 
  \scalebox{0.6}{ \includegraphics{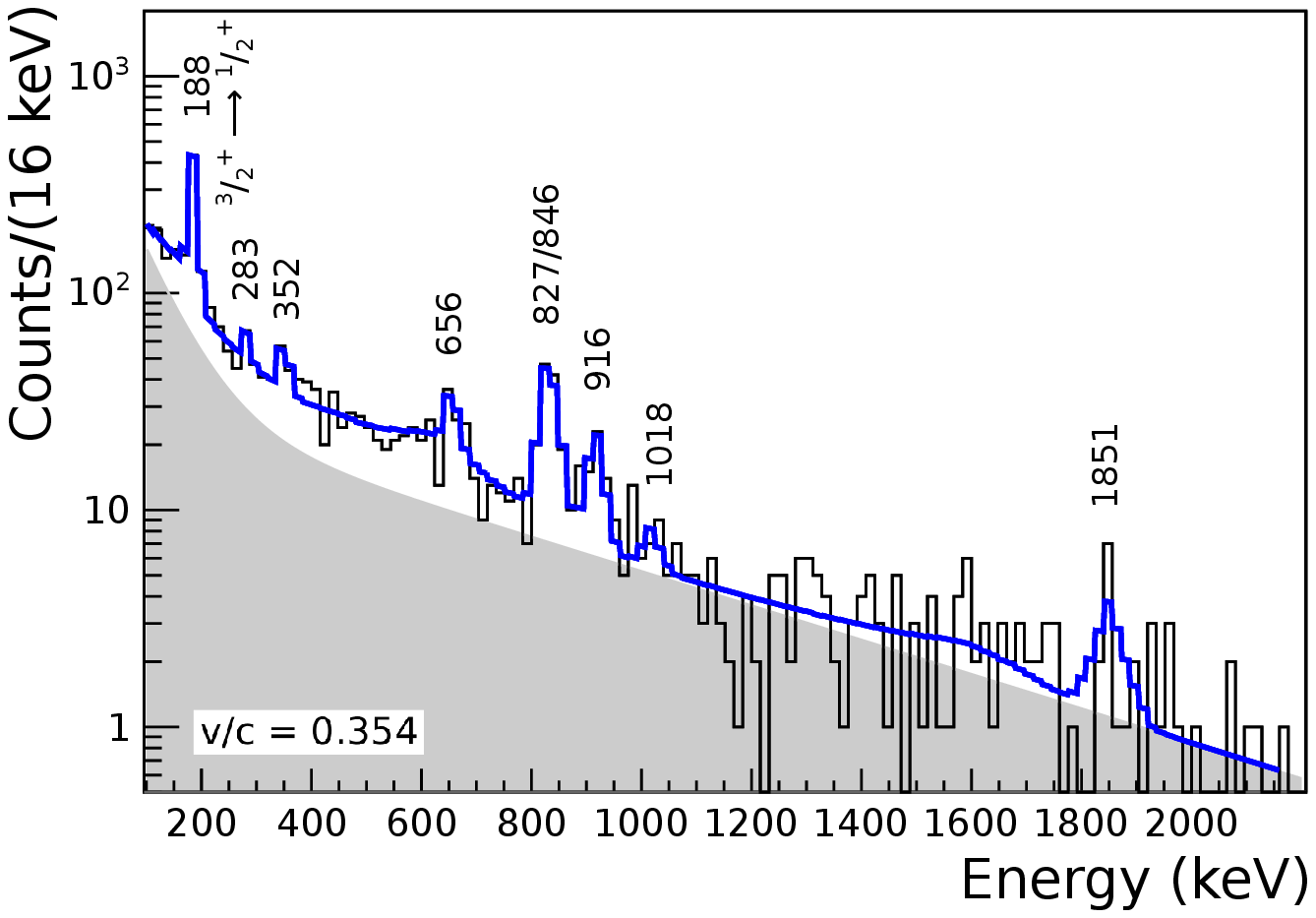} }
  \caption{\label{fig:p43_spectrum} (Color online) Projectile-frame
    spectrum of $^{43}$P measured via inverse-kinematics proton
    scattering.}
\end{figure}

Projectile-frame $\gamma$-ray spectra measured via inverse-kinematics
proton scattering from $^{42,44}$S and $^{41,43}$P appear in
Figs.~\ref{fig:s42_spectrum}-\ref{fig:p43_spectrum}. The average
projectile velocities in Table~\ref{tab:beams} were used in the Doppler
reconstruction of the $\gamma$ rays emitted in flight.  The solid
curves in the figures are fits consisting of a linear combination of
\textsc{geant4} simulations of the response of GRETINA to the observed
$\gamma$ rays with a prompt background, shaded in grey, consisting of
two exponential functions. The contribution of the non-prompt room
background to the fits was negligible. The $\gamma$-ray energies and
intensities extracted from the fits are listed in
Table~\ref{tab:gammas}. The $\gamma$-ray energies reported to the
right in Table~\ref{tab:gammas} were determined by varying the
simulated energies of the emitted $\gamma$ rays to optimize the fits
of the response functions to the measured spectra. The error ranges
correspond to 95\% confidence intervals.

\begin{figure} 
  \scalebox{0.6}{ \includegraphics{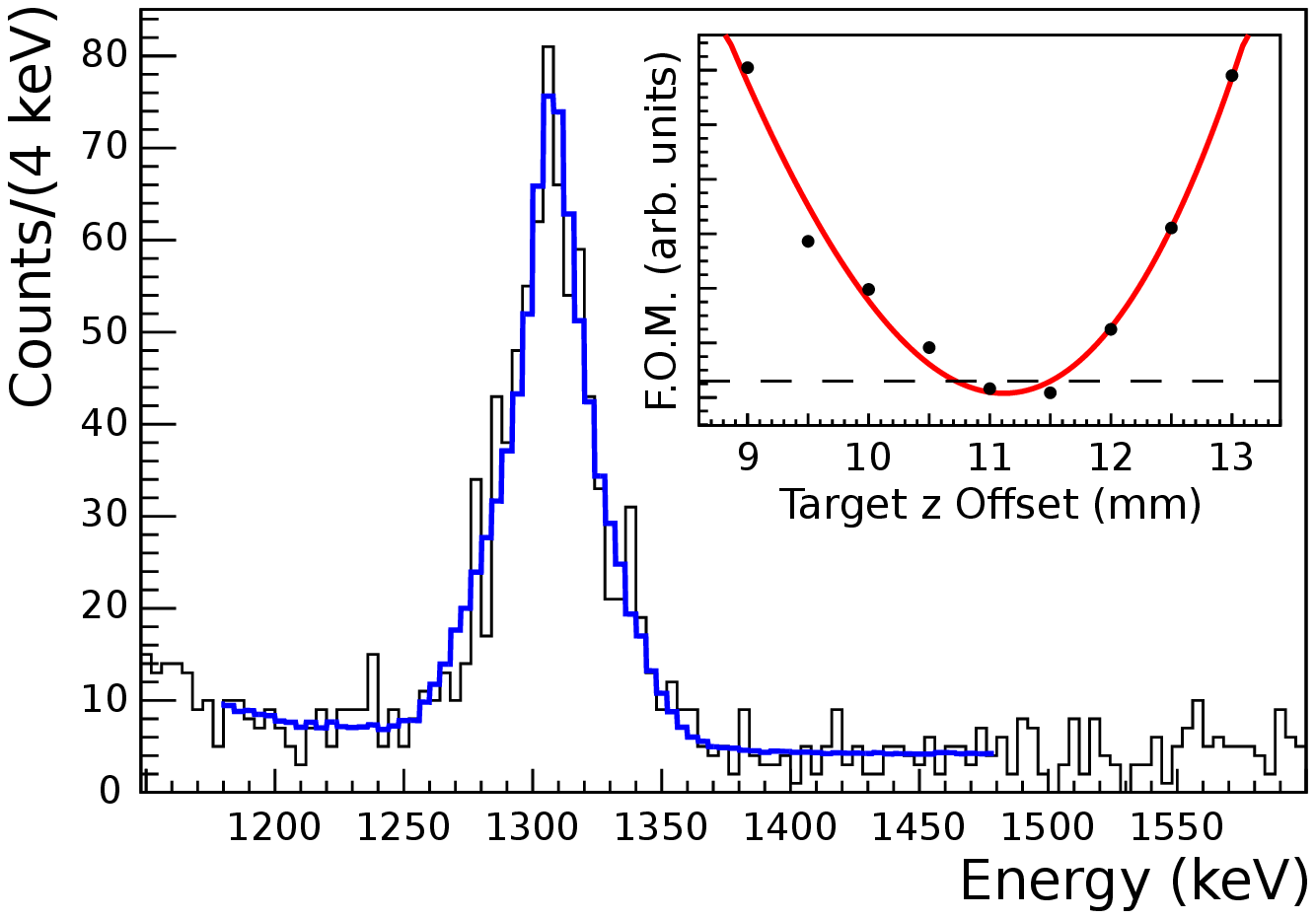} }
  \caption{\label{fig:z} (Color online)
    The region of the projectile-frame spectrum of $^{44}$S
    surrounding the photopeak of the 1329~keV $\gamma$ ray de-exciting
    the $2^+_1$ state.  The smooth curve is the \textsc{geant4} fit
    corresponding to the best-fit value of the target offset from the
    focus of GRETINA along the beam axis of 11.1~mm. The inset
    shows the figure of merit from the fit vs. the simulated target
    offset. The dashed line corresponds to the 95\% confidence
    interval of 0.4~mm.}
\end{figure}

The position of the liquid hydrogen target relative to the focus of
GRETINA along the beam axis strongly impacts the energies of $\gamma$
rays in Doppler reconstruction. We found the target offset to be
11.1(4)~mm by fixing the energy of the 1329~keV
$2^+_1 \rightarrow 0^+_\mathrm{g.s.}$ transition in $^{44}$S and
varying the target offset in simulations to obtain a best fit to the
measured spectrum. We chose this transition, because its energy was
determined to a precision of 1~keV in a measurement of $^{44}$S nuclei
at rest in the laboratory~\cite{For10}.  The best fit to the 1329~keV
peak along with a plot of the figure of merit from the log-likelihood
fit vs. the offset of the target along the beam axis appear in
Fig.~\ref{fig:z}.  We accounted for the mean lifetime of the $2^+_1$
state of $^{44}$S, deduced from the $B(E2;0^+_{g.s}\rightarrow 2^+)$
value measured via Coulomb excitation~\cite{Gl97}, of 3.5(10)~ps in
the simulations. However, we found that the impact of the lifetime
on the resulting best-fit target offset was below 0.1~mm, a
statistically insignificant effect relative to the 0.4~mm uncertainty
in the result.

\begin{figure} 
  \scalebox{0.6}{ \includegraphics{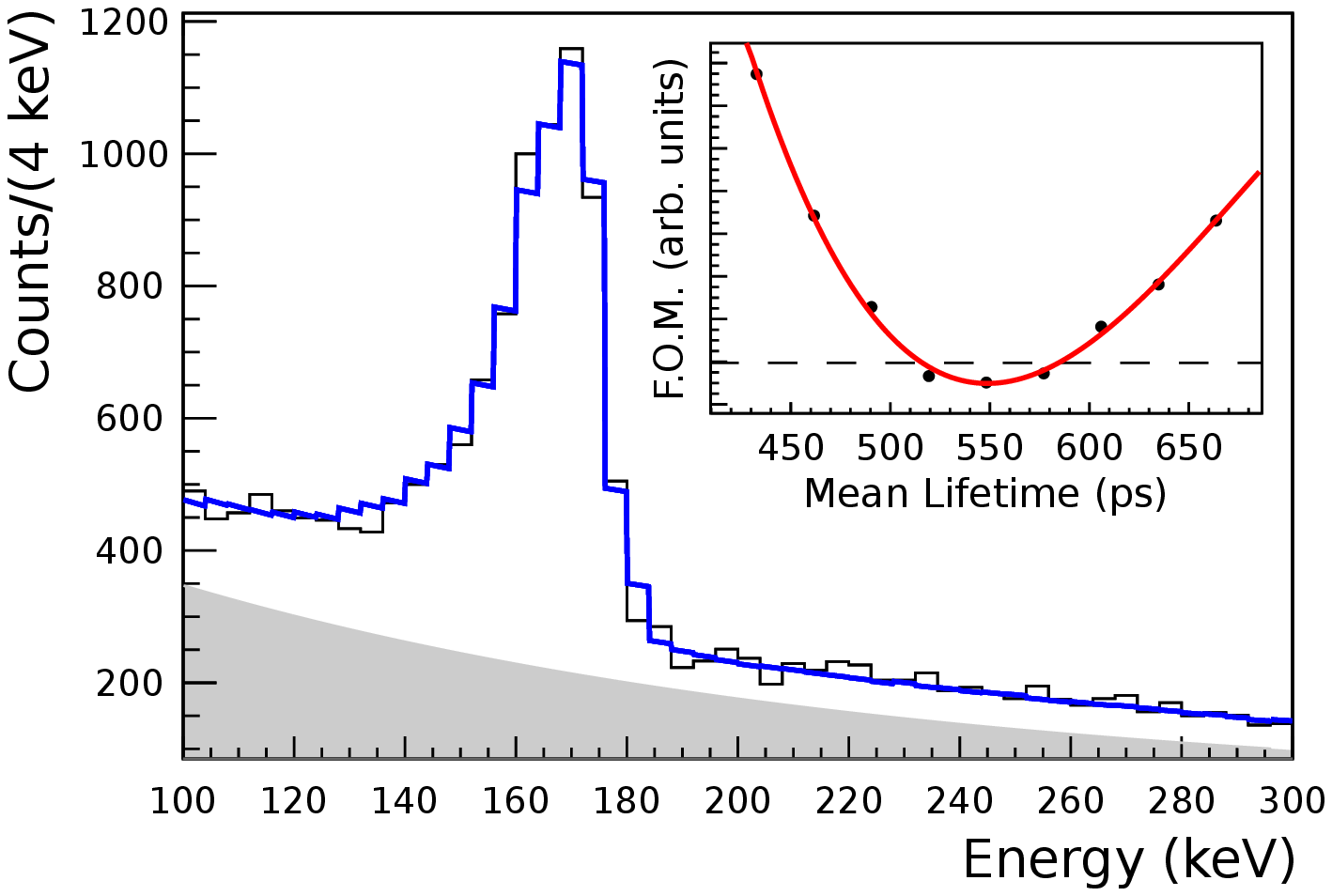} }
  \caption{\label{fig:p41_lt} (Color online)
    The low-energy region of the projectile-frame spectrum of $^{41}$P
    measured via inverse-kinematics proton scattering. The smooth
    curve is the \textsc{geant4} fit corresponding to a mean lifetime
    of the $J^\pi = (3/2^+)$ first excited state of 550~ps.
    The inset shows the figure of merit from the fit vs. the simulated 
    mean lifetime. The dashed line corresponds to the 95\% confidence
    interval of $\pm 35$~ps.}
\end{figure}

\begin{figure} 
  \scalebox{0.6}{ \includegraphics{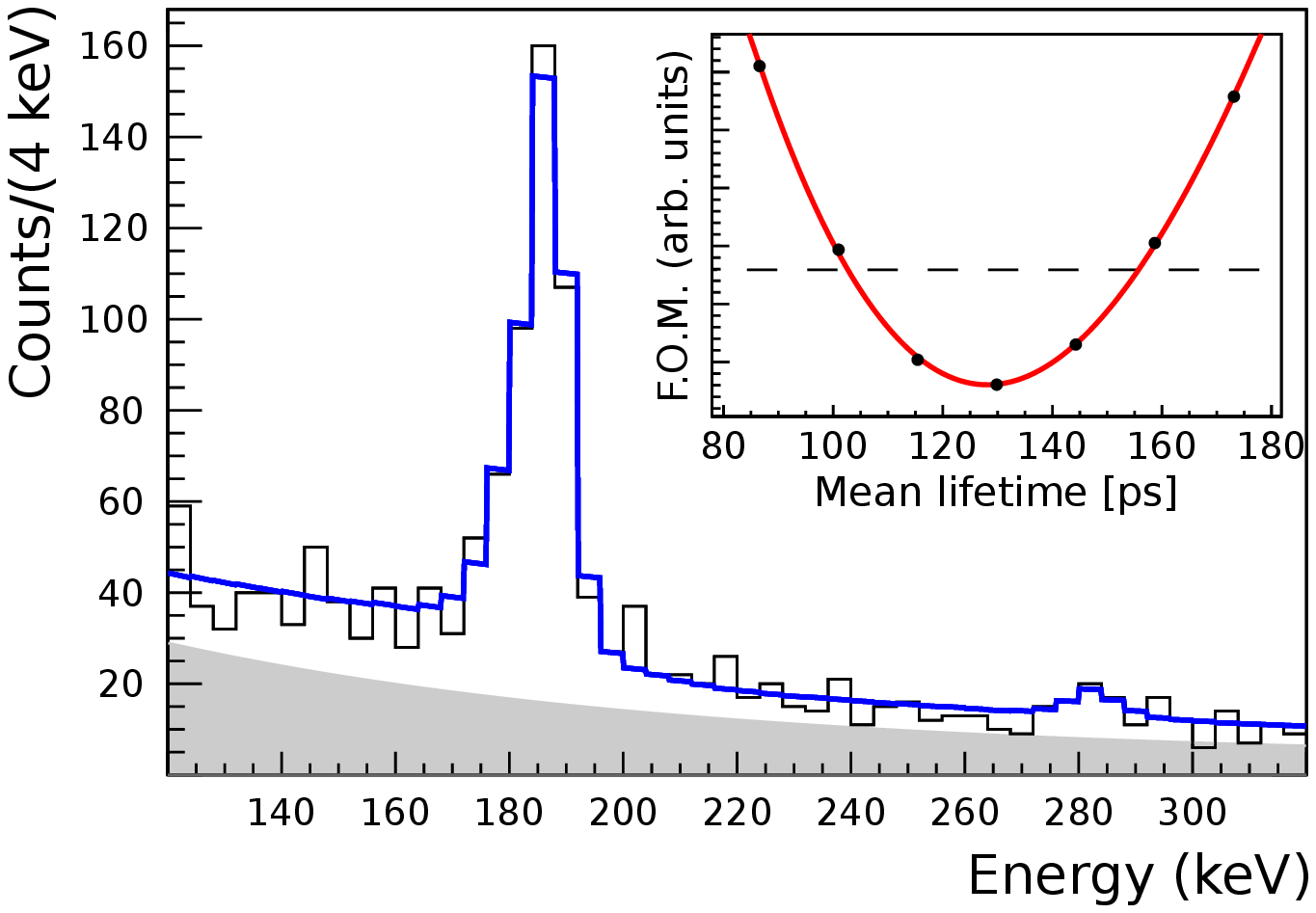} }
  \caption{\label{fig:p43_lt} (Color online)
    The low-energy region of the projectile-frame spectrum of $^{43}$P
    measured via inverse-kinematics proton scattering. The smooth
    curve is the \textsc{geant4} fit corresponding to a mean lifetime
    of the $J^\pi = 3/2^+$ first excited state of 130~ps.
    The inset shows the figure of merit from the fit vs. the simulated
    mean lifetime. The dashed line corresponds to the 95\% confidence
    interval of $\pm 30$~ps.}
\end{figure}

The mean lifetimes of the $2^+_1$ state of $^{42}$S of 20.6(15)~ps
and the $4^+$ state of $^{44}$S at 2457~keV of 76(24)~ps reported by
Parker \textit{et al.}~\cite{Par17} are long enough to impact the
Doppler-corrected $\gamma$-ray line shapes and were included in the
simulations of the 903~keV and 1150~keV $\gamma$ rays de-exciting
these states. These transitions were not observed with sufficient
statistics in the present work to perform independent mean lifetime
measurements.
The line shapes corresponding to the $\gamma$ rays
de-exciting the 173~keV and 188~keV first excited states of $^{41}$P
and $^{43}$P show low-energy tails consistent with 
lifetimes on the order tens to hundreds of picoseconds.
The low-energy regions of the
projectile-frame $\gamma$-ray spectra of $^{41}$P and $^{43}$P are
shown in Figs.~\ref{fig:p41_lt} and~\ref{fig:p43_lt}.
We varied both the energies and mean lifetimes of the states to
determine best-fit values of $\tau_{(3/2^+)} = 550(70)$~ps in $^{41}$P
and $\tau_{3/2^+} = 130(40)$~ps in $^{43}$P.
In both cases, the best-fit transition energy varies by less than
1~keV over the full uncertainty range of the mean lifetimes.  Plots of
the figure of merit from the log-likelihood fits vs. mean lifetime
appear in the insets in Figs.~\ref{fig:p41_lt} and~\ref{fig:p43_lt}.
In addition to the statistical uncertainties, we have included the
contribution of the 0.4~mm uncertainty in the position of the target
along the beam axis in the error ranges. This is an 11\% effect in
$^{41}$P and a 20\% effect in $^{43}$P.

\begin{table*}
\caption{\label{tab:gammas} Level energies, spins and parities,
  and $\gamma$-ray energies from
  Refs.~\cite{Lun16, Che11, Bas07, Ril08} and $\gamma$-ray energies,
  relative intensities, and cross sections from the present work.}
\begin{ruledtabular}
\begin{tabular}{crcrcrc}
&$E_\mathrm{level}$ [keV] & $J^\pi$ [$\hbar$] & $E_\gamma$ [keV] &
  $E_\gamma$ [keV] & $I_\gamma$ [\%] & $\sigma$ [mb]
\\\hline\hline
$^{42}$S    & \multicolumn{3}{c}{Ref.~\cite{Lun16}} &
&& \\\cline{2-4}
& 902      & $2^+$       &  902(4)  &  903(2)  & 100(9) & 23(6)\\
&2722      & $(4^+)$     & 1820(4)  & 1820(30) &   8(4) &  2.4(12)\\
&3002      & $(2^+)$     & 3002(4)  & 2990(30) &  20(5) &  6.2(15)\\
&          &             & 2100(4)  &          &   $<1$ &\\\cline{5-7}
&          &             & ---      & 1570(30) &   9(4) &\\
&          &             & ---      & 2190(30) &   8(4) &\\
$^{44}$S   &&\multicolumn{2}{c}{Ref.~\cite{Che11}} & &  \\\cline{3-4}
&1329      & $2^+$       & 1329.0(5) & 1329     & 100(4) & 15(3)\\
&2283(4)   & $(2^+)$     &  949(5)   & 954(4)   &  17(3) &  4.5(8)\\
&2479(11)  & $(4^+)$     & 1128(6)   & 1150(11) &  11(3) &  2.7(8)\\
&3264(6)   & $(2^+)$     & 1891(10)  & 1899(6)  &  13(2) &  3.7(7)\\
&          &             & 1929(7)   & 1955(25) &   2(2) &\\
&4027(13)  &             & ---       & 2698(13) &   8(2) &  2.1(5)\\\cline{5-7}
&          &             & ---       & 3076(10) &   8(2) &\\
$^{41}$P   &&\multicolumn{2}{c}{Ref.~\cite{Bas07}} & &  \\\cline{3-4}
& 173(1)   & $(3/2^+)$   &  172(12)  &  173(1)  & 100(2) &  6(2)\\
&1150(3)   &             &  964(22)  &  972(1)  &  43(2) & 13.5(6)\\
&          &             & 1146(28)  & 1148(2)  &  24(2) &\\
&1589(4)   &             & 1408(19)  & 1415(3)  &  23(2) &  4.6(4)\\
&          &             &  420(22)  &          &  $<2$  &\\
&2324(6)   &             & ---       &  733(5)  &   6(1) & 1.1(3)\\\cline{5-7}
&          &             & ---       & 1729(5)  &   9(1) &\\
$^{43}$P   && \multicolumn{2}{c}{Ref.~\cite{Ril08}} & & \\\cline{3-4}
& 188(1)   &  $3/2^+$    &  184(1)   &  188(1)  & 100(6) & 4(2)\\
& 845(6 )  & $(5/2^+)$   &  845(4)   &  846(11) &  15(7) & 4.9(14)\\
&          &             &  661(4)   &  656(6)  &  15(5) & \\
&1015(4)   & $(5/2^+)$   &  825(5)   &  827(4)  &  38(8) & 6.2(12)\\
&1104(5)   & $(5/2^+)$   &  911(6)   &  916(5)  &  22(5) & 3.5(8)\\
&2039      & $(5/2^+)$   &  1851(11) & 1851     &   7(4) & 1.8(8)\\
&          &             &  1018(6)  &          &  $<8$  &\\\cline{5-7} 
&          &             & ---       &  283(6)  &  7(3)  &\\
&          &             & ---       &  352(13) & 11(4)  &\\
\end{tabular}
\end{ruledtabular}
\end{table*}

\begin{figure*}
  \begin{tabular}{ccc}
  \scalebox{0.35}{
    \includegraphics{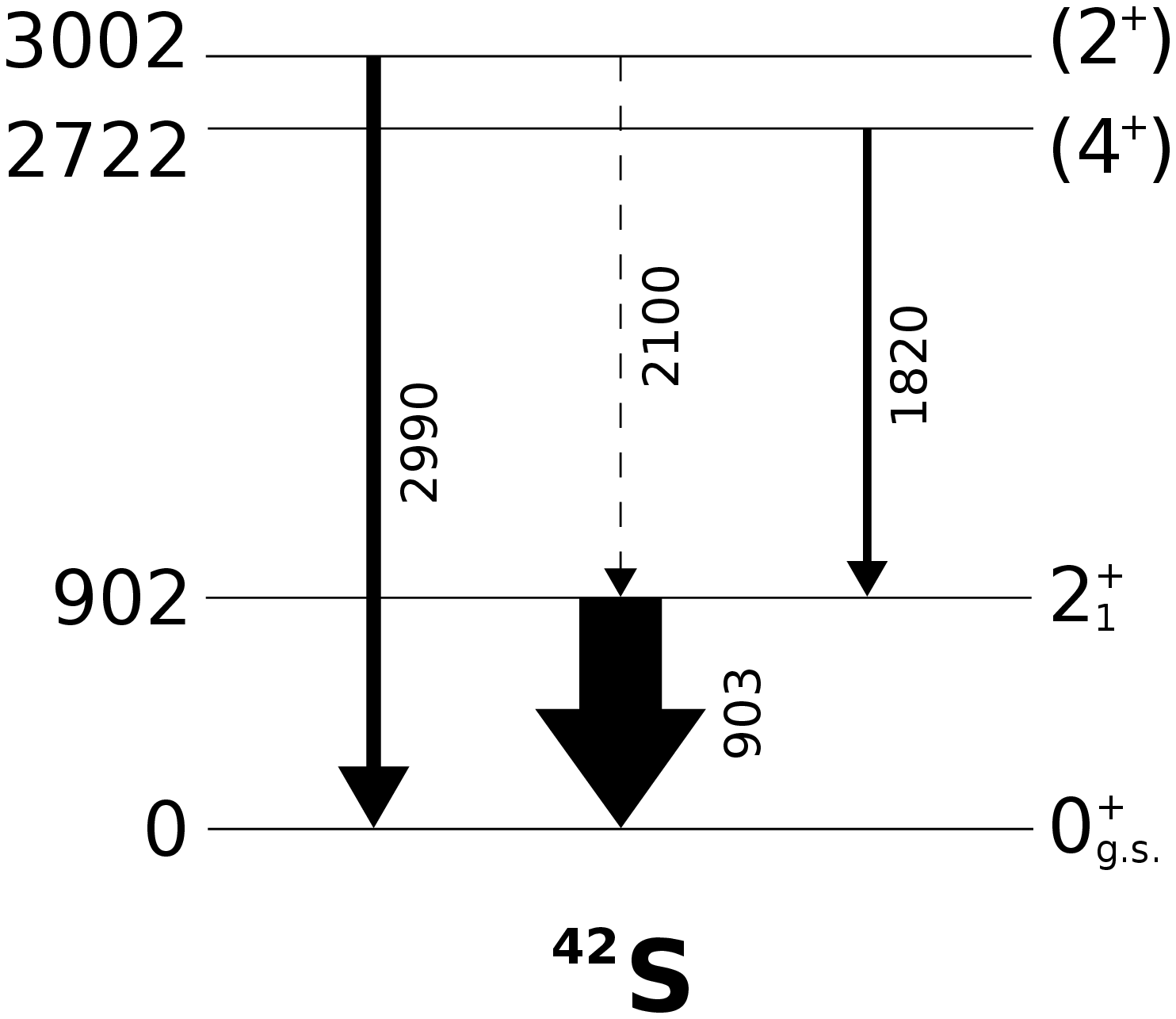}
  }
  & \hspace{1cm} &
  \scalebox{0.35}{
    \includegraphics{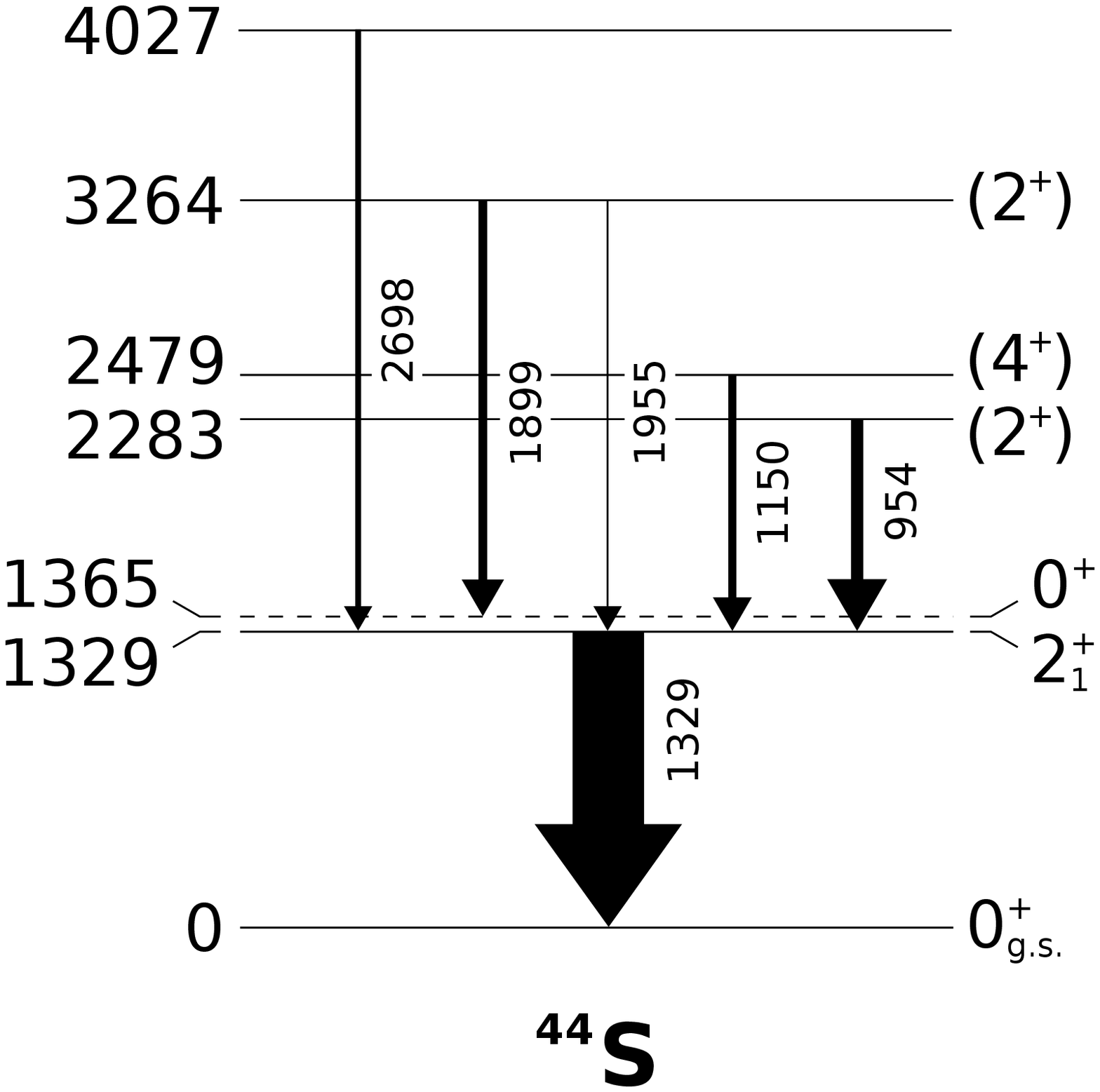}
  }\\
  &\\
  \scalebox{0.35}{
    \includegraphics{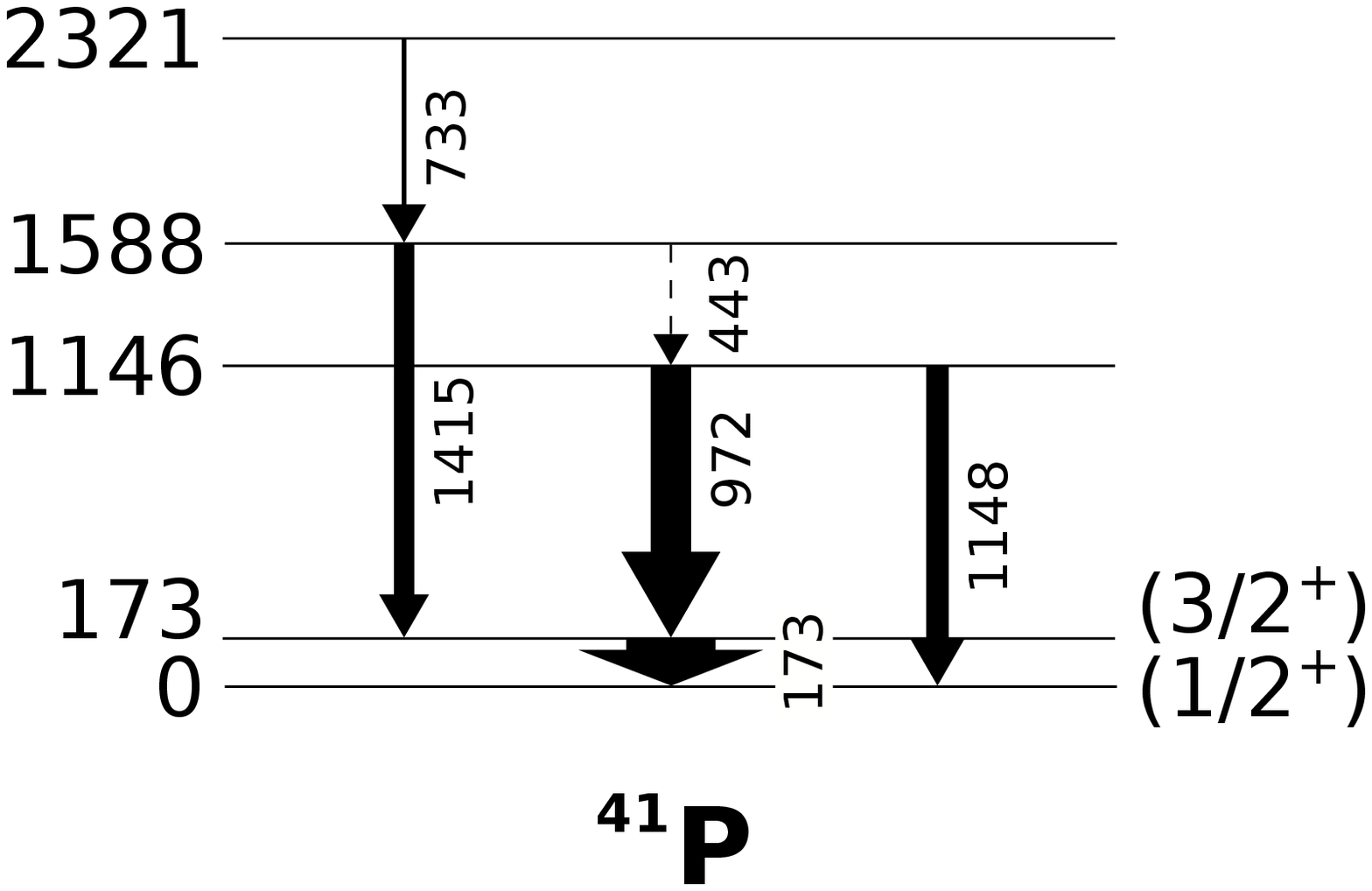}
  }
  &&
  \scalebox{0.35}{
    \includegraphics{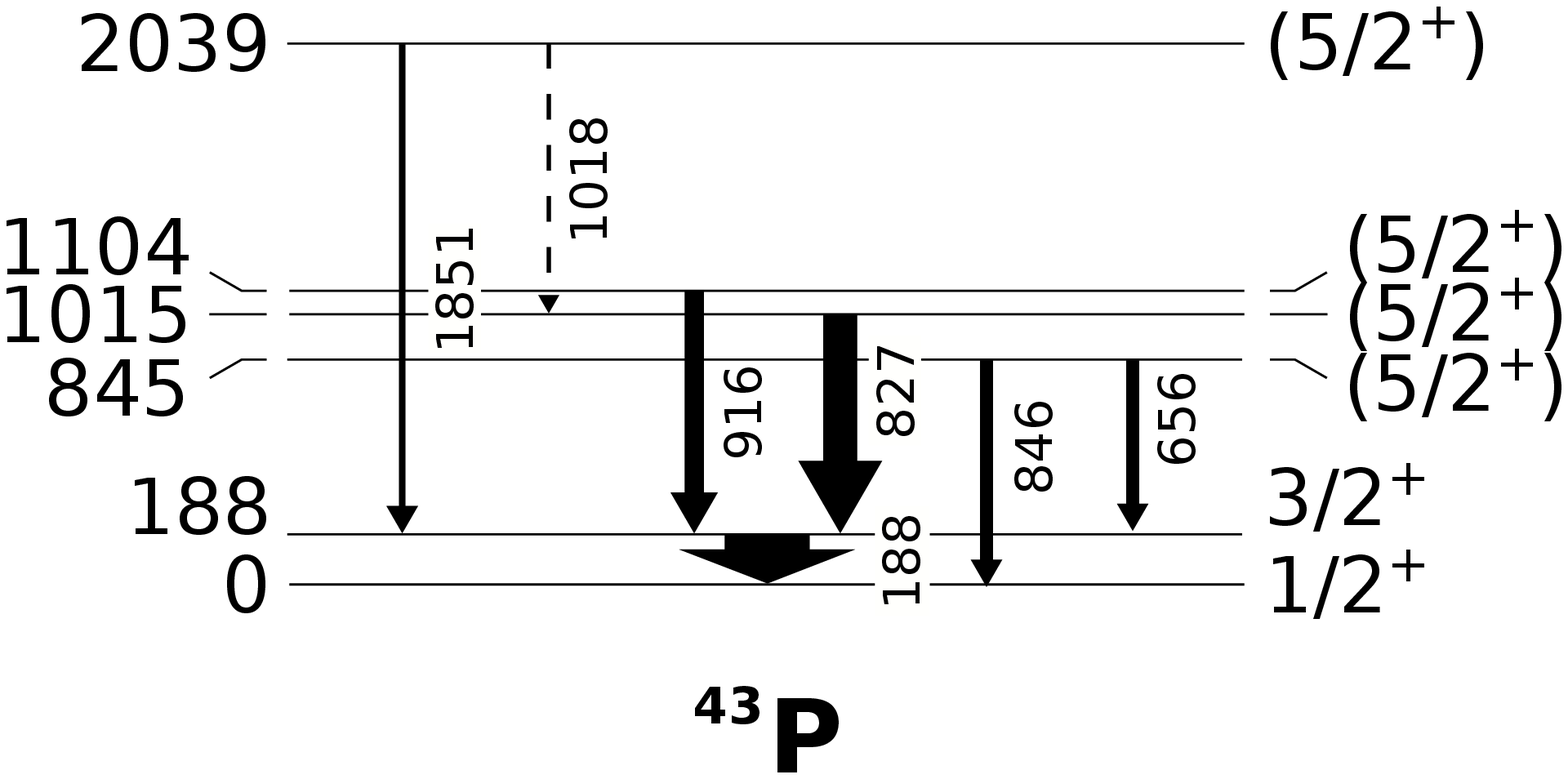}
  }
  \end{tabular}
  \caption{\label{fig:levels}  Partial level schemes of $^{42,44}$S
    and $^{41,43}$P showing levels populated in the present
    work. Arrow widths are proportional to the measured $\gamma$-ray
    intensities.}
\end{figure*}

Partial level schemes of $^{42,44}$S and $^{41,43}$P including the
levels populated in the present work are shown in
Fig.~\ref{fig:levels}.  We observe several known $\gamma$ rays, and
have identified two new transitions in $^{42}$S at 1570~keV and
2190~keV, which we are unable to place in the level scheme. In
$^{44}$S, we observe two new $\gamma$ rays at 2696~keV and
3076~keV. We place the 2696~keV transition feeding the $2^+_1$ state
due to the fact that it is seen in the spectrum of $\gamma$ rays
measured in coincidence with the 1329~keV $2^+_1 \rightarrow
0^+_\mathrm{g.s.}$ transition shown in
Fig.~\ref{fig:s44_spectra}(b). We are unable to place the 3076~keV
transition in the level scheme. In $^{41}$P, we place a new 733~keV
transition in the level scheme on the basis of its observation in
coincidence with the 1415~keV-gated spectrum in
Fig.~\ref{fig:p41_spectrum}(b), where we also see a possible weak
$\gamma$ ray at $\approx$930 keV, which we are unable to confirm or
place in the level scheme. The 420~keV transition observed by Bastin
\textit{et al.}~\cite{Bas07} to de-excite the state at 1588~keV along
with the 1415~keV $\gamma$ ray, and which appears in
Fig.~\ref{fig:levels} as a dashed arrow, was below our detection
threshold. We are also unable to place a new 1729~keV transition. In
$^{43}$P, we included in the fit the 1018~keV and 1851~keV transitions
de-exciting the excited state at 2035~keV observed in one-proton
knockout from $^{44}$S~\cite{Ril08}. We were only able to place an
upper limit on the intensity of the 1018~keV transition. We were
unable to place new 283~keV and 352~keV transitions.

The cross sections for inelastic proton scattering to excited states
of $^{42,44}$S and $^{41,43}$P listed in Table~\ref{tab:gammas} were
determined from the measured $\gamma$-ray yields, corrected for
feeding by transitions from higher-lying states, based on the partial
level schemes in Fig.~\ref{fig:levels}. In the case of the $2^+_1$
state of $^{42}$S, the 2100~keV $\gamma$ ray observed to de-excite the
3002~keV $(2^+)$ state by Lunderberg \textit{et al.}~\cite{Lun16} and
shown as a dashed arrow in Fig.~\ref{fig:levels} was below our
detection threshold. We included the 2100~keV $\gamma$ ray in the fit
to place an upper limit on its intensity, and included that upper
limit in the feeding correction.  We observed $\gamma$ rays, at
1570~keV and 2190~keV in $^{42}$S and at 3076~keV in $^{44}$S, that we
could not place in the respective level schemes. We have included
possible feeding of the $2^+_1$ states by these $\gamma$ rays in the
error ranges of the measured cross sections. We find cross sections for
populating the $2^+_1$ excitations via proton scattering of 23(6)~mb
in $^{42}$S and 15(3)~mb in $^{44}$S.

\begin{figure}
  \scalebox{0.6}{
    \includegraphics{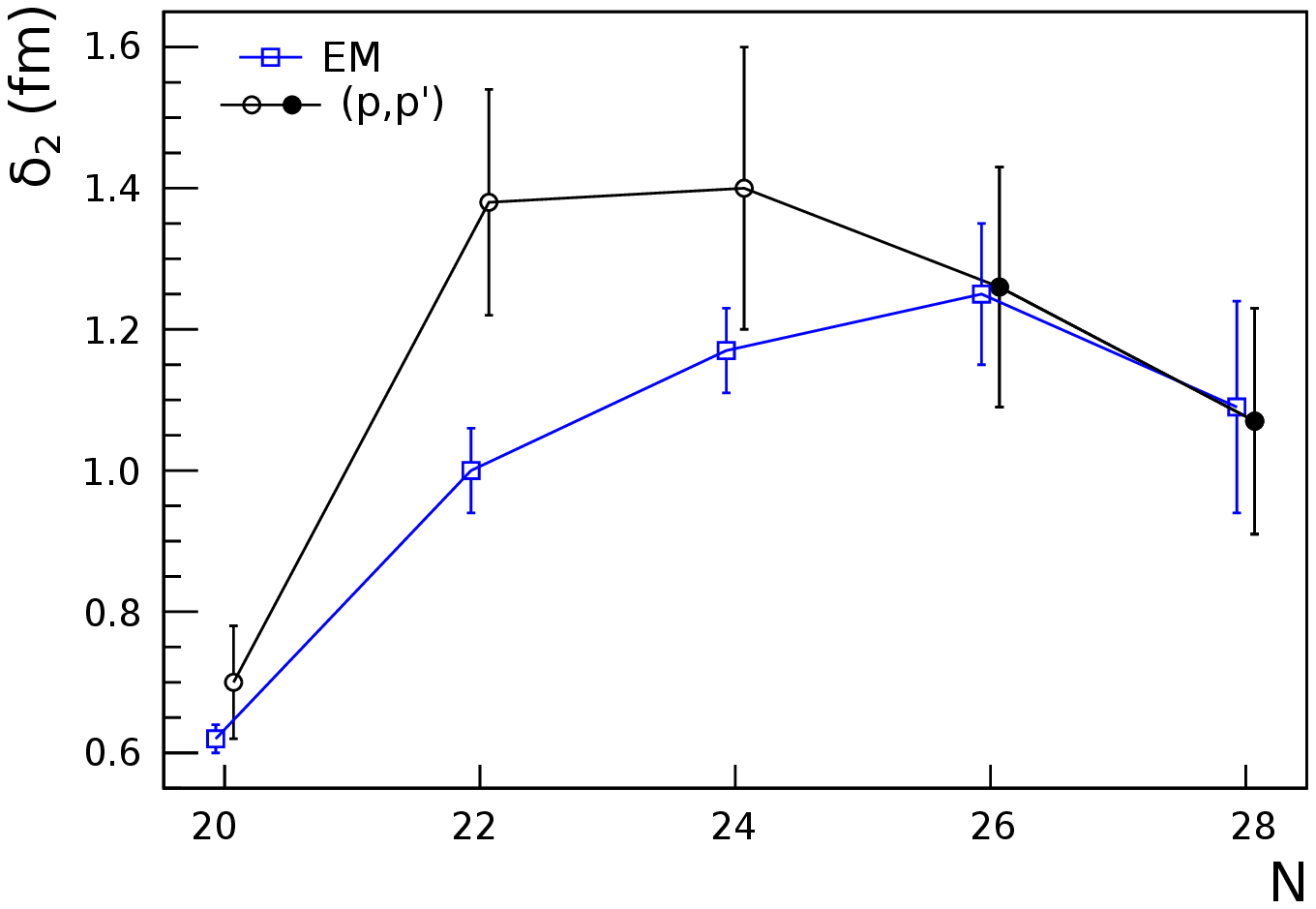}
  }
  \caption{\label{fig:delta2} (Color online) Proton-scattering and
    electromagnetic deformation lengths $\delta_2$ for the $0^+_{g.s.}
    \rightarrow 2^+_1$ excitations of even-even neutron-rich sulfur 
    isotopes. Proton-scattering deformation lengths are from 
    Ref.~\cite{Mar99} (open circles) and the present work (filled
    circles), and electromagnetic deformation lengths (open squares)
    are from Ref.~\cite{Pri14}.}
\end{figure}

We used the coupled-channels code \textsc{ECIS97}~\cite{Ray81,
  *ECIS97} and the global optical potentials of Ref.~\cite{Kon03}
and~\cite{Bec69} to determine deformation lengths from our measured
cross sections for inelastic scattering to the $2^+_1$ states of
$\delta_2 = 1.26(17)$~fm for $^{42}$S and $1.07(16)$~fm for
$^{44}$S. The error ranges include both the uncertainties in measured
cross sections and any discrepancy due to the two global optical
potential sets and the use of vibrational and rotational models. The
impact of the choice of optical potential parameter set was 2\% for
$^{42}$S and 7\% in the case of $^{44}$S. The variation in the 
deformation lengths determined using the vibrational and rotational
models for the excitations was at the 3\% level. Proton-scattering
deformation lengths of the $2^+$ states of even-even neutron-rich
sulfur isotopes from the present work and Ref.~\cite{Mar99} are
plotted along with electromagnetic deformation lengths from the
evaluation of Ref.~\cite{Pri14} in Fig.~\ref{fig:delta2}.

We did not collect sufficient statistics to perform $\gamma$-ray
angular distribution measurements. However, significant alignment of
the residual nucleus can be expected in direct reactions with fast
beams~\cite{Oll03, Stu03, Tak14}. We used the ECIS calculations
described above to evaluate the potential impact of $\gamma$-ray
angular distributions on our measured $\gamma$-ray yields. We
integrated the angular distributions from ECIS of the components
$t_{20}$ and $t_{40}$ of the polarization tensor of the $^{42,44}$S
nuclei after excitation to their $2^+_1$ states via $(p,p')$ to
determine their expectation values, which correspond to the
orientation parameters $B_2$ and $B_4$ in the usual notation. We found
roughly 30\% oblate alignment for $^{42}$S and 20\% oblate alignment
for $^{44}$S, with very similar $\gamma$-ray angular distributions
predicted by the vibrational and rotational models.  Following the
formalism outlined in Refs.~\cite{Oll03, Stu03}, we calculated the
corresponding angular distribution coefficients and performed
simulations including the resulting $\gamma$-ray angular distributions
and compared the resulting $\gamma$-ray yields with those obtained
assuming isotropic $\gamma$-ray emission in the projectile frame.  The
predicted angular distribution affected the yield of the 903~keV
$\gamma$ ray in $^{42}$S at the 3\% level and that of the 1329~keV
$\gamma$ ray in $^{44}$S at the 1\% level. These effects fall well
within the statistical uncertainties in the $\gamma$-ray yields. It is
important to note that the significant feeding of the $2^+_1$ states
by de-excitations of higher-lying states leads to a reduced degree
of alignment relative to these estimates.

\section{Discussion}

Given the importance of $^{42}$Si and neighboring isotopes for
building our understanding of nuclear structure close to the neutron
dripline, a range of observables in these nuclei should be measured
and understood.  In this section, we use the present
$^{42,44}$S$(p,p')$ results to extract $M_n/M_p$, the ratio of the
neutron and proton transition matrix elements for the
$0^+_{gs} \rightarrow 2^+_1$ excitations, which provide insights about
the presence or absence of closed shells.  In addition, we use the
$^{41,43}$P$(p,p')$ results to add to a comparison of the SDPF-U and
SDPF-MU shell model interactions recently begun by Gade \textit{et
  al.} in a study of $^{42}$Si~\cite{Gad19}.

The comparison of the present $^{42,44}$S$(p,p')$ results on the
$0^+_{gs} \rightarrow 2^+_1$ excitations with the previous Coulomb
excitation measurements of the same transitions \cite{Sc96,Gl97} allows
us to determine $M_n/M_p$.  Coulomb excitation measures the proton
transition matrix element exclusively, while proton scattering
involves both the proton and neutron transition matrix elements.  If
the excitation is isoscalar, then the ratio $M_n/M_p$ of the neutron
and proton transition matrix elements is equal to the ratio $N/Z$ of
the neutron and proton numbers, and hence $(M_n/M_p)/(N/Z)$ is equal
to 1.  This ratio is determined from the proton inelastic scattering
deformation length $\delta_{(p,p')}$ and the proton deformation length
$\delta_p$ using the equation,~\cite{Ber81}
\begin{equation}
\label{eq:MnMp}
\frac{M_n}{M_p} = \frac{b_p}{b_n} 
\left( \frac{\delta_{(p,p')}}{\delta_p}
\left( 1 + \frac{b_n}{b_p} \frac{N}{Z} \right) - 1 \right),
\end{equation}
where $b_n/b_p$ is the ratio of the sensitivities of the proton
scattering reaction to the neutron and proton contributions to the
excitation. The ratio $b_n/b_p$ is approximately 3 at proton energies
below 50 MeV, and approximately 1 at 1 GeV.  However, there is
considerable uncertainty about the value of $b_n/b_p$ at the energy of this 
experiment -- 60-70 MeV in the center of mass frame.  So for
the purposes of the present analysis, we assume that $b_n/b_p = 2 \pm
1$, which despite the large uncertainty allows us to reach important
conclusions about the present measurements. 

\begin{figure}
  \scalebox{0.6}{
    \includegraphics{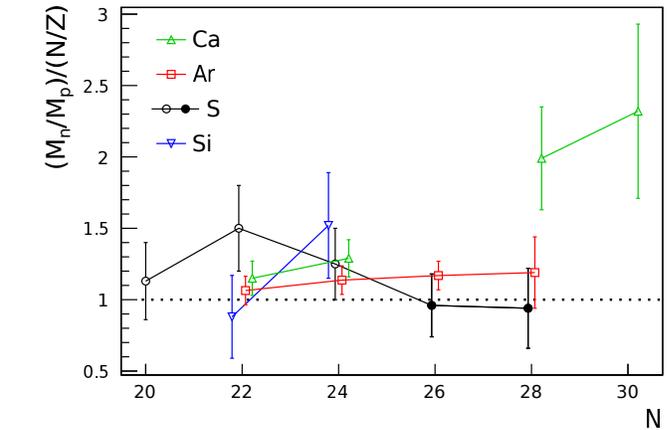}
  }
  \caption{\label{fig:MnMp} (Color online) Ratios of neutron to proton
    transition matrix elements $M_n/M_p$ expressed relative to $N/Z$
    for even-even neutron-rich calcium, argon, sulfur, and silicon
    isotopes from Refs.~\cite{Ber83,Ril14,Sch00,Ril05,Mar99,Cam07}
    (open symbols) and the present work (filled circles).}
\end{figure}

Figure~\ref{fig:MnMp} illustrates the values of $(M_n/M_p)/(N/Z)$ for
the $0^+_{gs} \rightarrow 2^+_1$ excitations in the $N=20-30$
even-even isotopes of Si, S, Ar and Ca.  We have combined
proton-scattering deformation lengths from the present work and
Refs.~\cite{Ber83,Ril14,Sch00,Ril05,Mar99,Cam07} with the proton
deformation lengths of Ref.~\cite{Pri14}.  Of the isotopes shown in
the plot, only two vary substantially from the value of
$(M_n/M_p)/(N/Z)=1$ expected for isoscalar transitions ---
the $N=28$ and 30 isotopes of Ca, which have a closed major proton
shell ($Z=20$). The $(M_n/M_p)/(N/Z)$ values for these two isotopes reflect 
the fact that while there are valence neutrons to contribute to
the $0^+_{gs} \rightarrow 2^+_1$ excitation, there are no valence
protons.  Therefore, the only proton contributions must result from the 
mechanism of core polarization. The results from the present work,
$(M_n/M_p)/(N/Z)=0.96(22)$ for $^{42}$S and $(M_n/M_p)/(N/Z)=0.94(28)$
for $^{44}$S are both statistically consistent
with 1.0.

\begin{figure}
  \scalebox{0.6}{
    \includegraphics{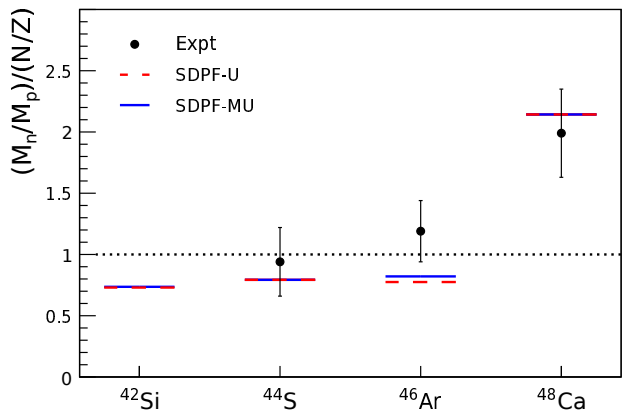}
  }
  \caption{\label{fig:MnMp28} (Color online) Ratios of neutron to
    proton transition matrix elements $M_n/M_p$ expressed relative to
    $N/Z$ for even-even $N=28$ isotones from Ref.~\cite{Ril14} and the
    present work compared with shell model predictions, described in
    the text.}
\end{figure}

In Fig.~\ref{fig:MnMp28}, the measured values for the $N=28$ isotones
are compared with shell model calculations performed with the SDPF-U
\cite{Now09} and SDPF-MU \cite{Uts12} effective interactions. For both
sets of shell model calculations, $M_p$ and $M_n$ are calculated using
the ``bare'' transition matrix elements $A_p$ and $A_n$ and a parameter
$d$ that reflects core polarization in the transitions: 
\begin{eqnarray}
\nonumber
M_p &=& A_p (1+d) + A_n (d)\\
M_n &=& A_p (d) + A_n (1+d)
\end{eqnarray}
We adopt $d=0.5$, which gives effective charges of $e_p = 1.5$ and
$e_n = 0.5$ for the electromagnetic transitions, values that were
used in Ref.~\cite{Gad19}. 

Calculations with both the SDPF-U and SDPF-MU interactions give 
identical or very similar $M_n/M_p$ values, with the largest
discrepancy of 6\% for $^{46}$Ar. The calculations predict that for
$^{42}$Si, $^{44}$S and $^{46}$Ar $M_n/M_p < N/Z$, signaling that
protons play a disproportionately large role in the $0^+_{gs}
\rightarrow 2^+_1$ excitations. The present $^{44}$S$(p,p')$ data are
not sufficient to distinguish between the isoscalar situation
($M_n/M_p=N/Z$) and the shell model predictions of $M_n/M_p=0.7$.

The shell model predictions for $Mn/Mp$ in $^{42}$Si, $^{44}$S and
$^{46}$Ar are provocative.  An $M_n/M_p$ value of less than $N/Z$
generally indicates a closed neutron shell.  Given the collapse of the
$N=28$ shell closure in $^{42}$Si and $^{44}$S, the shell model
predictions for $M_n/M_p$ in those nuclei are interesting and
important to test experimentally.

To determine whether the $M_n/M_p$ values for $^{42}$Si, $^{44}$S and
$^{46}$Ar are less than $N/Z$ and consistent with the shell model
predictions with statistical confidence, three issues will have to be
addressed.  First, the uncertainty in the $(p,p')$ data for these
nuclei will have to be reduced by increasing the numbers of counts in
the experimental spectra significantly -- and the new Facility for Rare
Isotope Beams (FRIB) will have that capability.  Second, a precise
Coulomb excitation measurement of $^{42}$Si is needed, and the
uncertainty in the Coulomb excitation result for $^{44}$S will need to
be improved, once again through the improvement in statistics possible
at FRIB.

Third, we must address the uncertainty in the ratio $b_n/b_p$, for
which we have used the value $2 \pm 1$.  Remarkably, this issue will
be addressed at FRIB as well.  The beams in the present work were at
energies of 70 MeV/nucleon and lower - energies for which there is
considerable uncertainty regarding the value of $b_n/b_p$ for inverse
kinematics $(p,p')$ reactions.  At FRIB, intense beams ($>10^4$
particles per second) of $^{42}$Si, $^{44}$S and $^{46}$Ar will be
available at energies much greater than 100 MeV/nucleon.  It has been
known for more than thirty years that inelastic hadron scattering at
energies over 100 MeV is approximately isoscalar; that is, it has
$b_n/b_p \approx 1$ (for example, see Ref. \cite{Kel88}).  Therefore,
performing inverse kinematics $(p,p')$ reactions at FRIB will nearly
eliminate the uncertainty in the value of $b_n/b_p$.

\begin{figure}
  \scalebox{0.6}{
    \includegraphics{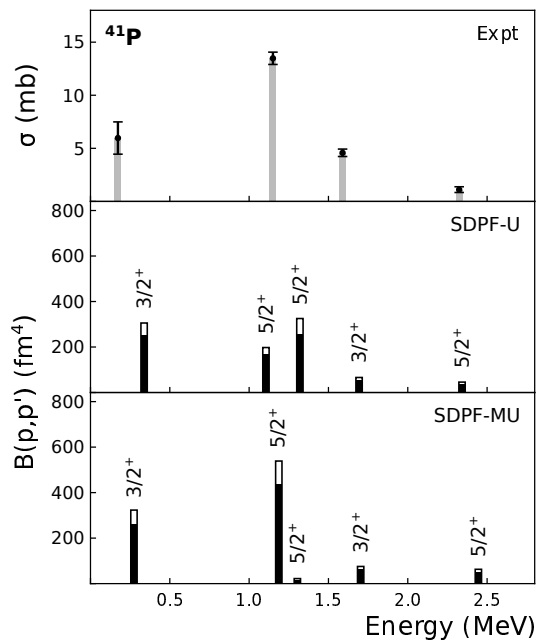}
  }
  \caption{\label{fig:p41_sm} Measured cross sections for populating
    excited states in $^{41}$P (top panel) compared with shell-model
    predictions of the proton-scattering transition strength
    $B(p,p')$ calculated using the SPDF-U~\cite{Now09} (middle panel)
    and SPDF-MU~\cite{Uts12} (bottom panel) effective interactions. In
    the bottom panels, the filled bars correspond to $b_n/b_p = 1$ and
    the open bars to $b_n/b_p = 3$.}
\end{figure}

\begin{figure}
  \scalebox{0.6}{
    \includegraphics{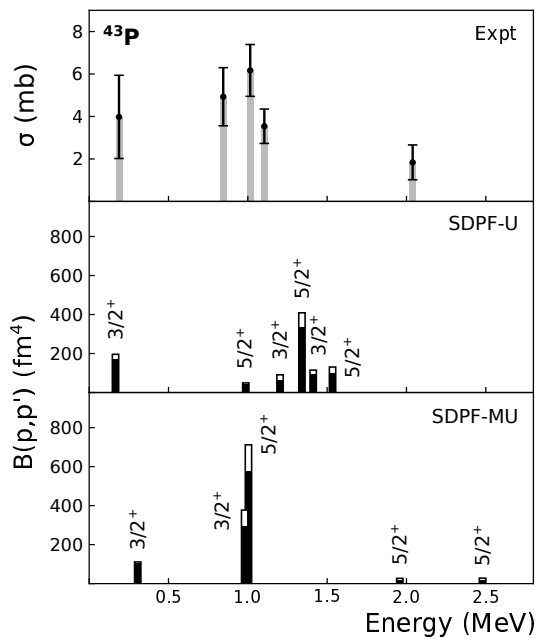}
  }
  \caption{\label{fig:p43_sm}  Measured cross sections for populating
    excited states in $^{43}$P (top panel) compared with shell-model
    predictions of the proton-scattering transition strength
    $B(p,p')$ calculated using the SPDF-U~\cite{Now09} (middle panel)
    and SPDF-MU~\cite{Uts12} (bottom panel) effective interactions.
    In the bottom panels, the filled bars correspond to $b_n/b_p = 1$
    and the open bars to $b_n/b_p = 3$.}
\end{figure}

The present $^{41,43}$P$(p,p')$ results provide an opportunity to
expand upon the recently reported comparison of the
SDPF-U~\cite{Now09} and SDPF-MU~\cite{Uts12} shell model interactions
using a measurement of the level scheme of $^{42}$Si via the
one-proton knockout reaction~\cite{Gad19}. The authors of
Ref.~\cite{Gad19} demonstrated that the SDPF-U interaction predicted a
number of states at low excitation energy (below 4~MeV) that was
significantly larger than what was observed in the experiment. In
contrast, SDPF-MU predicted a smaller number of states in the same
range of excitation energy that more accurately reflected the observed
spectrum. The authors of the $^{42}$Si study therefore concluded that
SDPF-MU is a more useful interaction for investigating the effects of
weak binding in $^{40}$Mg. Here, we find that the SDPF-MU is also
better able to describe the more deeply bound P isotopes in the
neighborhood of $^{42}$Si.

In Fig.~\ref{fig:p41_sm}, we compare the cross sections of the
states observed here in $^{41}$P$(p,p')$ with the distribution of
$(p,p')$ strength, $B(p,p')$, predicted using the SPDF-U and SPDF-MU
interactions. The strength $B(p,p')$ is calculated for each state
using the proton ($M_p$) and neutron ($M_n$) transition matrix
elements for the decay from the state to the ground state using the
equation,
\begin{equation}
B(p,p') = \frac{1}{(2J_i+1)} (C_p M_p + C_n M_n)^2
\end{equation}
where $J_i = 1/2$, since the ground states of both $^{41,43}$P have
$J=1/2$.  $M_n$ and $M_p$ are calculated for $^{41,43}$P in 
the same way that the corresponding values are calculated 
for the even-even isotopes (as described in the discussion above). 
The normalized coefficients $C_p$ and $C_n$ account for the 
sensitivity of $(p,p')$ to protons and neutrons and
take on the values $C_p = C_n = 0.5$ for $b_n/b_p = 1$ and $C_p = 0.25$,
$C_n = 0.75$ for $b_n/b_p = 3$. In the bottom two panels of
Fig.~\ref{fig:p41_sm} the filled bars correspond to $b_n/b_p = 1$, and
the open bars correspond to $b_n/b_p = 3$.  The values between 
the top of the filled bar and the top of the open bar correspond to the 
range $b_n/b_p=2 \pm 1$.

Figure~\ref{fig:p41_sm} shows that both the SDPF-U and SDPF-MU
interactions reproduce the experimental observation of the strong
excitation of the first excited state in $^{41}$P, which has
$J^\pi = 3/2^+$. SDPF-U gives two strong $5/2^+$ states at about
1.1~MeV, while SDPF-MU gives only one. The experiment shows only one
strong state at that energy, so that observation favors the SDPF-MU
interaction. The experiment shows a state near 1.6~MeV and another
near 2.4~MeV. Both SDPF-U and SDPF-MU give such states -- with a
$3/2^+$ state near 1.6~MeV and a $5/2^+$ state near 2.4~MeV.

We conclude that the comparison of the theoretical calculations with
the data on $^{41}$P favors SDPF-MU. 

Figure~\ref{fig:p43_sm} illustrates the situation in $^{43}$P. As in
$^{41}$P, the experiment shows that the lowest excited state is
strongly excited, and that excitation is reproduced by both the SDPF-U
and SDPF-MU interactions. In the experiment, there is a cluster of
three strongly populated states near 1~MeV. While SDPF-MU predicts two
strongly populated states near 1~MeV (one having $J^\pi = 3/2^+$ and the
other $J^\pi = 5/2^+$), the SDPF-U interaction gives a cluster of five
states distributed from 1.0 to 1.5~MeV, with the strongest being a
$5/2^+$ state near 1.4~MeV. We conclude that the SDPF-MU interaction
gives a better accounting of the situation in $^{43}$P than SDPF-U does. 

In short, the present results on $^{41,43}$P$(p,p')$ and the recently
reported $^{42}$Si results all favor the SDPF-MU interaction in this
neutron-rich region. 

\section{Conclusions}

Comparison of the proton inelastic scattering measurement of the
$0^+_{gs} \rightarrow 2^+_1$ excitation in $^{44}$S reported here with
previous Coulomb excitation measurements of the same excitation gives
an $M_n/M_p$ value that is, because of experimental uncertainties,
consistent with both the isoscalar value of $N/Z$ and the shell model
prediction that $M_n/M_p=0.7(N/Z)$, which would indicate that proton
excitations play a disproportionately large role in this excitation
and that there is a residual $N=28$ shell closure effect.  However,
the higher beam rates and higher beam energies available at FRIB will
provide an opportunity to resolve this issue not only in $^{44}$S but
also in $^{42}$Si.  The $^{41,43}$P$(p,p')$ measurements reported here
provide a means for expanding upon the comparison of the SDPF-U and
SDPF-MU shell model interactions begun in a recent study of $^{42}$Si
\cite{Gad19}. As in the case of $^{42}$Si, the present results favor
the SDPF-MU interaction.

\begin{acknowledgments}
This work was supported by the National Science Foundation under Grant
Nos.  PHY-1617250, PHY-1064819, PHY-1565546, and 1401574, and by the
U.S. Department of Energy under Grant DE-SC0009883. 
GRETINA was funded by the DOE, Office of Science. Operation of the
array at NSCL was supported by the DOE under Grant No. DE-SC0014537
(NSCL) and DE-AC02-05CH11231 (LBNL). 
We also thank T.J. Carroll for the use of the Ursinus College Parallel
Computing Cluster, supported by NSF grant no. PHY-1607335.
\end{acknowledgments}


%

\end{document}